# OPTICAL PROPERTIES OF $ZnP_2$ NANOPARTICLES IN ZEOLITE

O.A.Yeshchenko [a,1], I.M.Dmitruk [a], S.V.Koryakov [a] and Yu.A.Barnakov [b]

[a] *Physics Department, National Taras Shevchenko Kyiv University,*
*6 Akademik Glushkov prosp., 03127 Kyiv, Ukraine*
[b] *Advanced Materials Research Insitite, University of New Orleans, New Orleans, LA 70148*

We report that for the first time the nanoparticles of II-V semiconductor ($ZnP_2$) were prepared and studied. $ZnP_2$ nanoparticles were prepared by incorporation into zeolite Na-X matrix. Absorption, diffuse reflection (DR) and photoluminescence (PL) spectra of the $ZnP_2$ nanoclusters incorporated into the supercages of zeolite Na-X were measured at the temperature 77 K. Five bands $B_1$-$B_5$ are observed in both the DR and PL spectra demonstrating the blue shift from the line of free exciton in bulk crystal. We attribute the $B_1$-$B_5$ bands to some stable nanoclusters with size less than the size of zeolite Na-X supercage. We observed Stokes shift of the PL bands from the respective absorption bands. The nonmonotonic character of its dependence on the cluster size can be explained as the result of competition of the Frank-Condon shift and the shift due to electronic relaxation.

## 1. Introduction

Many different methods have been used for preparation of the semiconductor nanoparticles. These are, e.g. preparation of nanoparticles in solutions [1], glasses [2] or polymers [3]. However, it is not easy to control the size distribution of small clusters with countable number of atoms and molecules in these methods. Matrix method based on the incorporation of materials into the 3D regular system of voids and channels of zeolites crystals could be one of the possible solution [4,5]. Zeolites are crystalline alumosilicates with cavities, size of which can vary in the range from one to several tents nanometers. It depends on the type of alumosilicate framework, ratio Si/Al, origin of ion-exchanged cations, which are stabilized negative charge of framework and etc. Zeolite NaX, which have been used in the present work has Si/Al ratio equal 1, Fd3m symmetry and two types of cages: one is sodalite cage – truncated octahedron with diameter 8 A and supercage, which formed by the connection of sodalites in diamond law with the diameter of about 13 Å [6]. All cages are interconnected by shared small windows and arranged regularly. Thus, the cages can be used for the preparing of small semiconductor clusters of zinc diphosphide ($ZnP_2$).

Many works have been reported on the nanoparticles of II-VI [1,7,8], III-V [9,10], IV-VI [11], I-VI [12], II-VII, IV-VII, V-VII [13] and IV [14] types of semiconductors. But to the best of our knowledge the nanoparticles of II-V type have not been studied yet. The present paper is the first study of $ZnP_2$ clusters. Wet chemistry methods are not applicable to scale up a production of ultrasmall $ZnP_2$ clusters due to their high reactivity at air. It is hard to expect its stability in glass melt as well. Thus, incorporation into zeolite cages seems to us one of the most suitable methods of preparation of $ZnP_2$ clusters.

Quantum confinement of charge carriers in small particles leads to new effects in optical properties of the particles. Those are the blue shift of exciton spectral lines originating from the increase of the kinetic energy of charge carriers and the increase of the oscillator strength per unit volume [15,16]. These effects are quite remarkable when the size of the particle is comparable with or smaller than Bohr radius of exciton in bulk crystal. Since the exciton Bohr radius in bulk monoclinic $ZnP_2$ ($\beta$-$ZnP_2$) is quite small (15 Å) [17], the problem to reach the quantum confinement for this material is quite difficult. Therefore, zeolite is the good candidate to solve this problem, as its cages are rather small and can contain only small clusters.

Bulk $\beta$-$ZnP_2$ crystal is the direct-gap semiconductor. As this crystal is strongly anisotropic, its optical spectra are characterised by three exciton series. The lowest energy exciton peak is observed at 1.55913 eV [18]. As the bulk crystal has rather small energy gap (1.6026 eV), the blue shifted exciton lines of $ZnP_2$ nanoparticles are expected to be in the visible spectral region.

## 2. Technological and experimental procedures

For the preparation of $ZnP_2$ nanoclusters we used crystals of $\beta$-$ZnP_2$ and synthetic zeolite of Na-X type. The size of zeolite single crystals was about 50 μm. Zinc diphosphide was of 99.999% purity. The framework of zeolite Na-X consists of sodalite cages and supercages with the inside dianeters of 8 and 13 Å, respectively. $ZnP_2$ molecule seems to us to be too large to be incorporated into small sodalite cage, because of the existence of many Na cations. Therefore, it is naturally to assume that only the supercages can be the hosts for $ZnP_2$ nanoparticles. Zeolite and $ZnP_2$ crystal were dehydrated in quartz ampoule in vacuum about $2 \cdot 10^{-5}$ mm Hg for one hour at 400°C. After that ampoule was sealed. We used 100-mm length ampoule for space separation of $ZnP_2$ source and zeolite in it. $ZnP_2$ was incorporated into the zeolite matrix through the vapour phase at 840°C in source region and 835°C in zeolite region for 100 hours. The cooling of ampoule we carried out gradually with inverted temperature gradient.

In our experiments we used the samples with 5 wt% loading level of $ZnP_2$ into zeolite. Since Samples in quartz ampoule were dipped into liquid nitrogen during the experiment. A tungsten-halogen incandescent lamp was used as a light source for the absorption and diffuse reflection measurements. An Ar$^+$ laser with wavelength 4880 Å was used for the excitation of the luminescence.

---

[1] Corresponding author. *E-mail address:* yes@univ.kiev.ua (O.A. Yeshchenko).

### 3. Results and discussion

Absorption, diffuse reflection (DR) and photoluminescence (PL) spectra of the ZnP$_2$ clusters incorporated into the 13 Å supercages of zeolite Na-X were measured at the temperature 77 K. The absorption and DR spectra of the studied nanoparticles are presented in fig.1. Conventional absorption spectrum was obtained from the transmission one measured from the layer of zeolite with ZnP$_2$ with the thickness of 0.3 mm. The absorption spectrum has no features. The absorption coefficient increases monotonically with increase of the photon energy in the range from 1.66 eV to 2.60 eV approximately. The absorption spectrum can be obtained from the diffuse reflection spectrum converting it with Kubelka-Munk function $K(\hbar\omega) = [1 - R(\hbar\omega)]^2 / 2R(\hbar\omega)$, where $R(\hbar\omega)$ is the diffuse reflectance normalised by unity at the region of no absorption. Obtained spectrum is more interesting than conventional absorption one as it demonstrates clear band structure. Five bands signed as $B_1$-$B_5$ is observed in the spectrum (fig. 1). The spectral positions of these bands are presented in table 1. All these bands demonstrate the blue shift (table 1) from the line of free exciton (1.55913 eV) in the bulk β-ZnP$_2$ crystal. The observed blue shift allows us to attribute these bands to the absorption of ZnP$_2$ nanoclusters incorporated into supercages of zeolite. It is often observed for nanoclusters that clusters with certain number of atoms are characterised by high binding energy (more stable clusters) and are more abundant in the sample. This effect is well known, e.g. for the nanoparticles of II-VI semiconductors [19,20]. We estimate that the largest ZnP$_2$ cluster in zeolite Na-X supercage with the diameter of 13 Å is (ZnP$_2$)$_7$. Under the assumption of the full loading of the supercages by nanoparticles for 5 wt% loading level of ZnP$_2$ into zeolite one can estimate the the average number of the ZnP$_2$ molecules per a supercage to be 5.7. Thus, $B_1$-$B_5$ bands can be attributed to these stable clusters containing up to 7 of ZnP$_2$ molecules. Probably they are stoichiometric since we do not have any evidence of ZnP$_2$ decomposition at the temperature of 840°C used for its sublimation. Results of theoretical search of the stable clusters in this size range will be printed in the next paper. The photoluminescence spectrum (fig. 2) shows the same structure as the DR spectrum, i.e. PL spectrum consists of the same five $B_1$-$B_5$ bands. Their spectral positions are presented in table 1. The PL bands are characterised by blue shift from the exciton line in bulk crystal as well (table 1).

The observed blue shift of the absorption and luminescence bands is the result of the quantum confinement of electrons and holes in ZnP$_2$ clusters. As the exciton Bohr radius $a$ in bulk crystal is 15 Å that is larger than radius of zeolite supercage $R = 6.5$ Å, the strong confinement regime takes place in the clusters. The experimental value of the blue shift obtained from the spectral positions of $B_5$ band in DR spectrum is 0.282 eV. In Ref. [21] the finite confining potential was used to calculate blue shift. However, for experimental blue shift of 0.282 eV the theory gives the value of cluster diameter of about 30 Å. That is considerably larger than the diameter of zeolite supercage (13 Å). So, the value of blue shift calculated by the effective mass approximation are substantially different from the experimental one. It means probably that the effective mass approximation fails for considered small clusters. One more explanation of the obtained small value of the blue shift is the shift of electron and hole energy levels to the low energy due to the tunnelling between the neibouring supercages.

As it can be seen from fig.1 the intensities of the bands increase at the increase of the respective photon energy in DR spectrum. Meanwhile, the opposite situation takes place for PL spectrum (fig.2). Here the intensities of bands increase at the decrease of the energy. This effect can be explained in such way. Lower intensities of the PL bands corresponding to the smaller clusters can be explained by reabsorption of their emission by larger clusters. The reabsorption is due to overlap of the corresponding bands, absorption into excited states and phonon-assisted transitions.

One can see from the table 1 that the luminescence bands have the Stokes shift from the absorption ones. The value of this shift is from 0.078 to 0.135 eV. Such shift is well known both in the molecular spectroscopy and in the spectroscopy of nanoparticles. It is known that this kind of Stokes shift (so-called Frank-Condon shift) is due to vibrational relaxation of the excited molecule or nanoparticle to the ground state. The dependence of the red shift in the spectra of $ZnP_2$ clusters in zeolite on the size of cluster is nonmonotonic. For smallest clusters the increase of the Stokes shift at the decrease of the nanoparticle size is observed. Such dependence can be explained by the theory developed, e.g. in Ref. [22] where the first-principle calculations of excited-state relaxations in nanoparticles and the dependence of respective Stokes shift on particle size were performed. As it is shown in Ref. [22], for smallest clusters the observed Stokes shift is the Frank-Condon shift which is the result of the structure relaxation of the cluster in excited and ground states. Meanwhile, for larger clusters the opposite dependence is observed, namely the Stokes shift increases with the increase of the cluster size. Perhaps, such behaviour is the result of the relaxation of initially excited electron-hole pair to its self-consistent equilibrium configuration with hole in the centre of cluster.

Table 1. Spectral characteristics of $ZnP_2$ nanoclusters in zeolite Na-X.

| Cluster size | Band | Spectral position (eV) | | Blue shift of absorption bands (eV) | Stokes shift (eV) |
| --- | --- | --- | --- | --- | --- |
| | | Absorption | PL | | |
| ↓ | $B_1$ | 2.367 | 2.268 | 0.808 | 0.099 |
| | $B_2$ | 2.200 | 2.115 | 0.641 | 0.085 |
| | $B_3$ | 2.068 | 1.995 | 0.509 | 0.078 |
| | $B_4$ | 1.955 | 1.854 | 0.396 | 0.101 |
| | $B_5$ | 1.842 | 1.706 | 0.282 | 0.1353 |

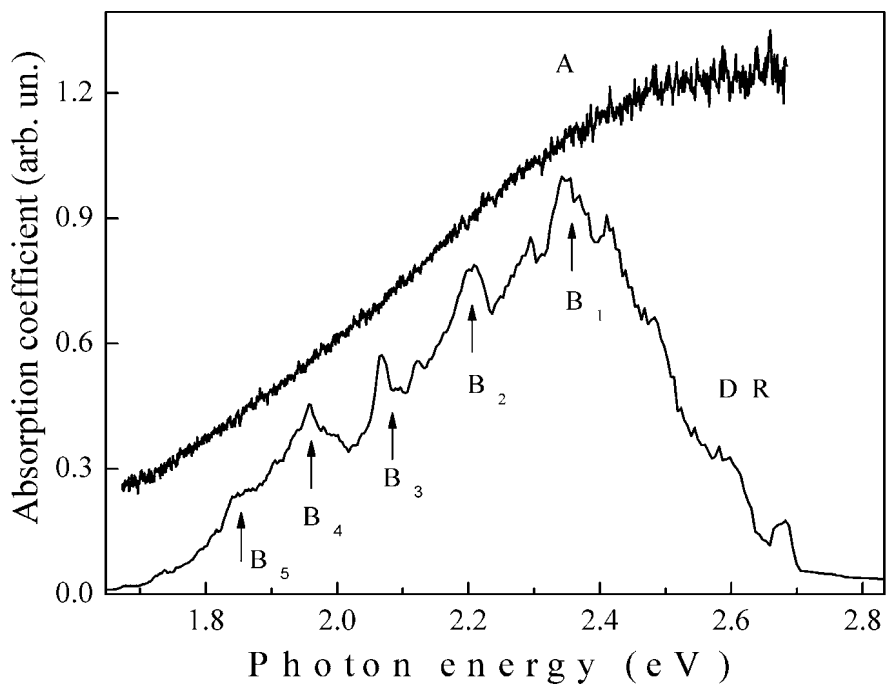

Fig. 1. The conventional absorption (A) and obtained by Kubelka-Munk method absorption spectra (DR) of ZnP$_2$ nanoclusters in zeolite Na-X at the temperature 77 K.

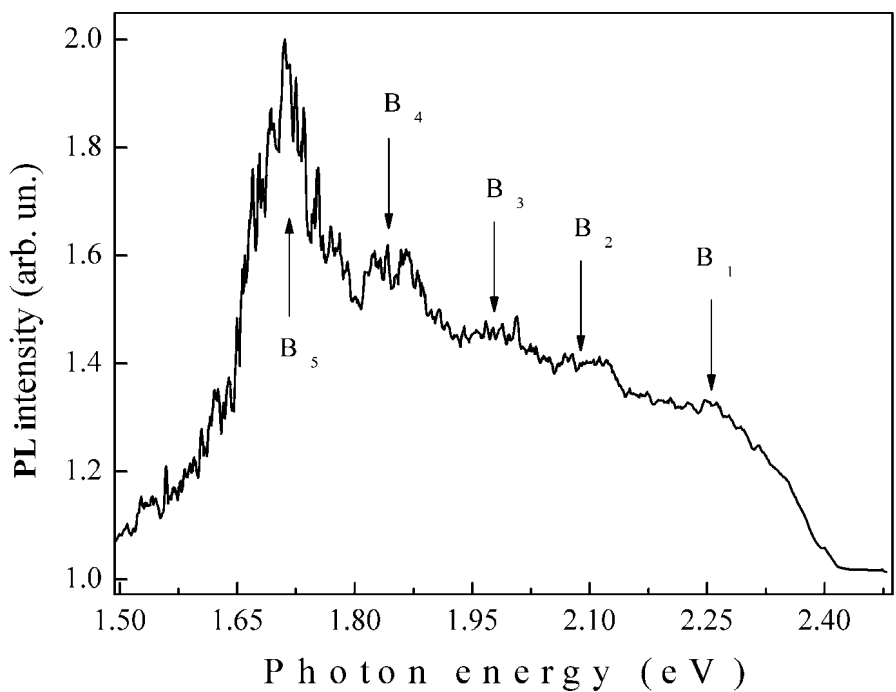

Fig. 2. The photoluminescence spectrum of ZnP$_2$ nanoclusters in zeolite Na-X at the temperature 77 K.